\documentstyle[12pt,epsfig]{report}
\begin{document}

\begin{center}

{\Large\bf
Chemical Equilibrium and Isotope\\
\vspace*{0.3cm}

Temperatures\footnote{
to appear in "Isospin Physics in Heavy-Ion Collisions
at Intermediate Energies"\\
Eds. Bao-An Li and W. U. Schr\"oder, Nova Science Publishers, Inc., in 
preparation} 
}
\vspace*{1.0cm}

J. Pochodzalla
\vspace{0.1cm}

{\it Max-Planck-Institut f\"ur Kernphysik,
D-69117 Heidelberg, Germany}
\vspace{0.2cm}

and 
\vspace{0.2cm}

W. Trautmann
\vspace{0.1cm}

{\it Gesellschaft f\"ur Schwerionenforschung mbH, 
D-64291 Darmstadt, Germany}

\vspace{1.0cm}

{\bf
ABSTRACT}
\end{center}
\vspace{0.3cm}

\noindent
The measurement and interpretation of isotopic yield ratios in heavy ion 
reactions at intermediate and high energies are discussed and the 
usefulness of these observables for establishing equilibrium properties 
and for determining thermodynamic parameters is illustrated.
The examples are mainly taken from work performed with lighter projectiles 
at intermediate and high energies and from studies of spectator reactions 
at relativistic energies. As an application, the caloric curve of nuclei,
as derived for Au on Au collisions, is introduced and discussed.
\vspace{1.0cm}

\newpage

\section{Introduction}
\label{Sec_1}

Isotopic yield ratios have proven to be useful observables
for studying the mechanisms of heavy-ion reactions at 
intermediate and high energies. 
Production yields for isotopically resolved 
particles and nuclear fragments and combinations thereof can provide us 
with answers to the questions of mutual stopping and subsequent 
equilibration of the collision partners.
To the extent that equilibrium is reached, they permit 
the extraction of the corresponding thermodynamical 
variables. The entropy, density, and particularly the temperature of the 
ensemble of excited nuclear systems, formed in the course of energetic 
encounters, have been deduced from measured isotopic yield ratios. 
These parameters provide a characterization of heavy-ion reactions 
whose complex dynamics advocate a global description in statistical terms. 
Temperature observables and their correlation with the excitation energy, 
commonly termed caloric curve of nuclei, offer 
the possibility to explore the link between the liquid-gas phase 
transition predicted for infinite nuclear matter and the decay properties 
of finite nuclei. 

The examples considered in this chapter are heavy-ion reactions 
at bombarding energies in the range of about 100 MeV per nucleon
up to a few GeV per nucleon.
The dynamics of these reactions, high above the Fermi energy, are 
predominantly governed by nucleon-nucleon collisions and by the 
possibility of the production (and reabsorption) of secondary particles, 
mostly pions. Nuclear matter properties are being probed at densities 
far away from the ground state density and at excitation energies up 
to the binding limit of nuclear matter and beyond. The corresponding
part of the phase diagram of extended nuclear matter includes the
spinodal and coexistence regions of the phase transition from the normal
liquid phase to a gas-like phase consisting of nucleons and light 
complex particles. Consequences of these nuclear phase properties 
are expected and searched for in this class of reactions.

The isotopic degrees of freedom in the nuclear disassembly and their
role in a statistical description will be the primary subject of this 
chapter. Isospin, therefore, will not appear here in its role as a 
symmetry principle of nuclear structure that follows from the charge 
independence of the nuclear forces. 
The third component of the isospin,
$t_3 = (N-Z)/2$,
is simply used to characterize isotopes according to
their neutron and proton numbers $N$ and $Z$.
Isospin equilibration, consequently, will indicate that isotopic 
yield distributions correspond to the expectations for chemical equilibrium. 
Isospin equilibration, in this sense, is a consequence of chemical 
equilibration and as such a necessary condition for equilibrium that can be 
tested experimentally.

The experimental
determination of the mass number $A$ of a reaction product is,
generally, more difficult than the determination of the element 
number $Z$ alone. It involves either improved resolutions, if the 
$\Delta E-E$ technique is used, or measurements of additional quantities, 
such as the
time-of-flight or the magnetic rigidity of the emitted reaction product.
Yield ratios of neighboring isotopes, in return, may be assumed to be
little influenced by dynamical effects such as Coulomb, recoil, or size 
effects, and thus should permit a fairly unbiased look at the balance
equation
\begin{equation}
(A+1,Z) \longleftrightarrow (A,Z) + n.
\label{EQ1}
\end{equation}
In a statistical description this balance is determined 
by the temperature 
and the chemical potentials, the measurement thus gives access to
these equilibrium parameters.

We will begin this chapter with a discussion of some of the earlier 
experiments on isotopic effects, performed with beams of fairly light 
ions of intermediate energy and of light particles at high energies
in the GeV range. 
These data will allow us to make the distinction between 
isospin mixing and isospin equilibration. We will then more generally 
discuss equilibrium in spectator reactions with heavier ions of incident 
energies up to one GeV per nucleon. The determination of 
equilibrium parameters, in particular of the temperature, and the 
construction of a caloric curve of nuclei will be described in the final
sections of this chapter. 

\section{Isospin Mixing and Equilibration}
\label{Sec_2}

One of the earlier experiments on isotopic degrees of freedom has
been conducted at the CERN synchrocyclotron
during the eighties of the last century \cite{wada87}. 
In this experiment, the isotopic composition of decay products was studied
as a function of the neutron-to-proton ($N/Z$) content of nuclear 
systems over the range accessible with stable projectile and target nuclei.
Targets of $^{58,64}$Ni, $^{\rm nat}$Ag, and $^{197}$Au were
bombarded with $^{12}$C and $^{18}$O beams with incident energy 84 MeV per 
nucleon. The produced particles and fragments were measured 
inclusively with high-resolution telescopes placed at several angles 
between $\theta_{\rm lab}$ = 40$^{\circ}$ and 120$^{\circ}$. 
Each telescope consisted of an axial-field ionization chamber followed 
by three silicon detectors of increasing thickness and a 
bismuth germanate scintillation detector. Full isotope 
separation for ions up to carbon was achieved for fragment energies 
$E/A \ge$ 3 MeV and for ions up to beryllium at somewhat lower energies.

Energy-integrated isotope yields were obtained by extrapolating from the 
isotopically resolved parts of the energy spectra with the help of fitting 
functions which were derived from the model of a Maxwellian source 
moving in beam direction. The measured isotope ratios were found to 
vary little with the 
detection angle. The following discussion is therefore based on the 
data measured at $\theta_{\rm lab}$ = 40$^{\circ}$ which have the highest 
statistics and the smallest experimental uncertainties. This angle 
is also sufficiently backward to be outside the range of projectile 
fragmentation.

\begin{figure}[t]
   \centerline{\epsfig{file=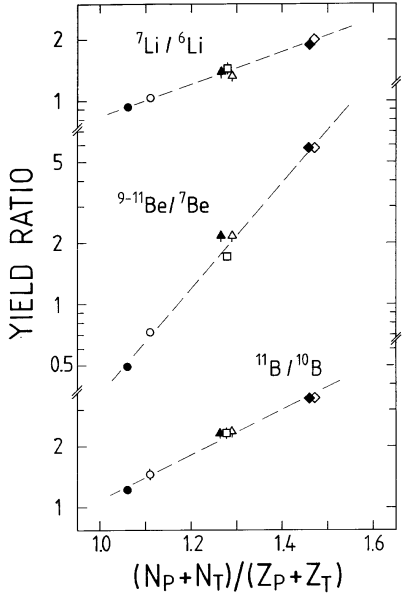,height=8cm}}
        \caption[]{\it\small
Ratios of energy-integrated isotope yields, measured at $\theta_{\rm lab}$ = 
40$^{\circ}$, as functions of the $N/Z$ ratio of the combined system of 
projectile and target. Closed and open symbols denote $^{12}$C and
$^{18}$O projectiles, respectively; circles, squares, triangles, and 
diamonds stand for $^{58,64}$Ni, $^{\rm nat}$Ag, and $^{197}$Au targets
(from Ref. \protect\cite{wada87}).
}
\label{rati}
\end{figure}

Three yield ratios, of lithium, beryllium, and boron isotopes and for all 
the investigated reactions, are shown in Fig. \ref{rati}, 
plotted as functions of 
the neutron-to-proton ratio $N/Z$ of the combined system of projectile 
and target nuclei. This way of representing the data was motivated by the 
following observations: The isotope-yield ratios vary strongly with the 
choice of the target. The relevant property of the target, however, is 
neither the mass number $A$ nor the atomic number $Z$ but rather the 
ratio $N/Z$; the 
yield ratios are different for the $^{58}$Ni and $^{64}$Ni targets but, 
apparently, the same for the $^{64}$Ni and $^{\rm nat}$Ag targets which have 
equal $N/Z$. Furthermore, the different $N/Z$ of the $^{12}$C and $^{18}$O
projectiles seem to have little influence except in the reactions with the 
$^{58}$Ni target. The fact that in this case the ratio measured with the 
$^{12}$C projectiles are significantly smaller than those measured with 
$^{18}$O suggests that the $N/Z$ of the combined system rather than that of 
the target nucleus alone is the relevant ordering parameter. Plotted in 
this way, all three isotope-yield ratios follow smooth exponential
functions and increase monotonically with $N/Z$ (Fig. \ref{rati}).

We may first conclude that the combined system of the projectile and the 
whole target is involved in the process of emission of complex 
fragments in these reactions. For the moment,
we will call this complete mixing, as expected for projectile nuclei that
are completely stopped in the target prior to the 
fragment emission. The two collision partners are thought to 
form a sufficiently 
homogeneous system such that the individual contributions of the projectile 
and target can no longer be separately identified. Even though confirmed
for a variety of reactions \cite{deak91,andro94} this is by no means 
a trivial result; there are many examples for which complete mixing is not 
observed: Isotope yield ratios of projectile fragments 
measured at small angles with $^{12}$C beams at the same energy 
84 MeV per nucleon are virtually constant for the same range of 
targets \cite{moug81}. 
Similar observations were made at GANIL where the fragmentation 
of heavier projectiles was studied \cite{guer83,lucas87,bord90}. 
For the set of reactions with $A$ = 40
projectiles and $A$ = 58 targets, discussed in detail by Li and Yennello
in Chapter 19, the stopping 
is found to become incomplete at incident energies of 45 MeV
per nucleon and above \cite{yenn94,john96,john97}.
For the mass symmetric reactions with $^{96}$Zr and $^{96}$Ru
projectiles and targets, studied at the much higher energy of
400 MeV per nucleon, the mixing was found to increase with the
centrality of the collision but to remain incomplete even at
the smallest impact parameters (Ref. \cite{rami00} and Chapter 19). 

\begin{figure}[t]
   \centerline{\epsfig{file=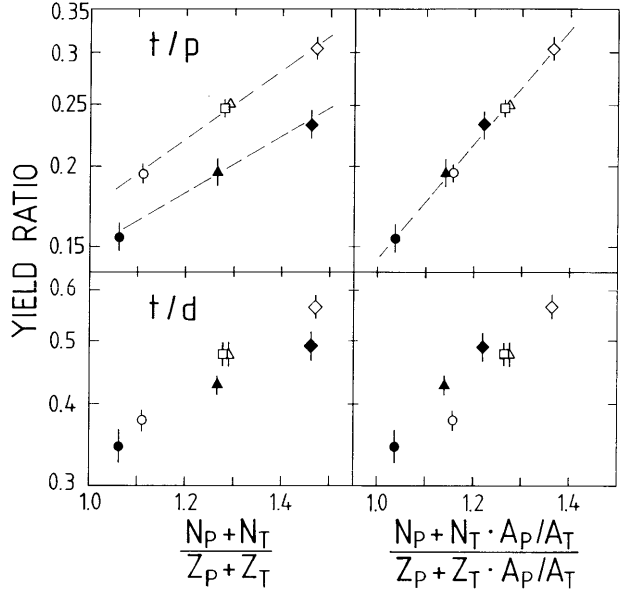,height=8.0cm}}
        \caption[]{\it\small
Ratios of preequilibrium 
triton-to-proton yields (top) and triton-to-deuteron yields
(bottom), measured at $\theta_{\rm lab}$ = 40$^{\circ}$ and plotted as 
functions of the $N/Z$ ratio of the combined system (left-hand side)
and of a source consisting of equal numbers of nucleons from the 
projectile and from the target (right-hand side). The symbols are chosen as 
in Fig. \ref{rati} (from Ref. \protect\cite{wada87}).
}
\label{lcpt}
\end{figure}

Incomplete mixing has even been observed in the same set of reactions 
at 84 MeV per nucleon with $^{12}$C and $^{18}$O projectiles, as 
illustrated in Fig.~\ref{lcpt}. There the ratios of hydrogen isotopes 
from the preequilibrium component measured at 
$\theta_{\rm lab}$ = 40$^{\circ}$ are shown for different mixing assumptions.
With the postulate that the observed ratios should be a unique function of 
the $N/Z$ ratio of the emitting source, the light particle data are 
inconsistent with emission from the combined system (Fig. \ref{lcpt}, 
left panels).
The t/p ratios can be reconciled with this postulate if the emitting source 
is assumed to consist of equal numbers of nucleons from the projectile and 
from the target, chosen according to their respective $N/Z$ ratios 
(Fig. \ref{lcpt}, upper right panel). 
For the t/d ratios the same assumption overcorrects the 
deviations (lower right panel) but by 
mixing projectile and target nucleons with the ratio 1:2 
a unique and monotonic increase 
of the measured isotope ratios can be achieved. These examples show that 
preequilibrium light particles originate from subsystems considerably
smaller than the combined system of projectile and target.

The question of isospin equilibration will now be addressed by asking for 
the significance of the slopes characterizing the dependence of the isotope 
ratios on the source $N/Z$. Their trends can be most easily understood by 
considering the grand canonical approach \cite{mekji78,albergo,barz88}. Here 
the yield ratios of neighboring isotopes can be expressed as
\begin{equation}
\frac{Y(A+1,Z)}{Y(A,Z)} = (\frac{A+1}{A})^{3/2} \cdot \frac{\omega (A+1,Z)}
{\omega (A,Z)} \cdot \exp(\frac{\mu_{\rm n} + \Delta B}{T})
\label{EQ2}
\end{equation}
where $\omega (A,Z)$ denotes the internal partition function and $\Delta B$
is the difference of the binding energies of the two nuclides.
The chemical potentials $\mu_{\rm n}$ of neutrons and $\mu_{\rm p}$ of protons
guarantee the conservation of the mean mass 
and charge of the disassembling nucleus within a given volume and,
therefore, are functions of the $N/Z$ ratio of the system. The assumption
that $\mu_{\rm n}$ depends, to first order, linearly on $N/Z$ immediately
explains the experimental finding 
that the logarithm of the ratio increases linearly with the 
$N/Z$ ratio of the emitting source. It further explains why the slopes 
are about the same for the pairs of lithium and boron isotopes which
both differ by one neutron (Fig. \ref{rati}).
The $^7$Be isotope is, at least, two neutrons lighter than the most 
abundant heavier isotope, and a term $2\cdot \mu_{\rm n} /T$ (or higher) 
appears in the exponent on the right-hand side of Eq. \ref{EQ2}. 
Consequently, the slopes should be about twice as large which is 
also in agreement with the observation. 
For this qualitative consideration, it has been tacitly assumed that 
the temperature $T$ is approximately the same for the considered 
fragment species and reactions, a condition that is furthermore 
needed for ensuring
constancy of the ratios of the internal partition functions.
It will be shown below that this assumption is 
not unrealistic in the present case (Section \ref{Sec_6}).

It is an instructive exercise to estimate the $N/Z$ dependence expected 
from other conceivable mechanisms \cite{traut87}.
In particular, the binomial distribution following from the assumption of a 
combinatorial mechanism of fragment formation 
yields a variation in proportion to $N/(N+Z)$ for ratios of neighboring
isotopes which is by far too weak to describe the data. 
The observed dramatic variation of the isotope 
ratios with $N/Z$ is a consequence of the exponential dependence 
on the ratio $\mu_{\rm n}/T$ and thus may by itself indicate
chemical equilibrium. 
Equilibrium in this sense does not necessarily imply complete mixing.
The equilibration may be local as in the examples of incomplete mixing 
discussed above where strong variations as a function of 
the neutron-to-proton ratio of the assumed localized 
sources were observed. Global equilibrium, clearly, requires
complete mixing.
 
\section{The Isotopic Effect}
\label{Sec_3}

An alternative possibility to study isospin equilibration is provided 
by the socalled isotopic effect \cite{boga74,lozh92}. It is generally 
defined as the net effect observed when switching to another target
made from a different isotope of the same element with no other change of
the experimental conditions. Here we will specifically consider
ratios of production cross sections of particular fragment species 
in otherwise identical reactions on the tin isotopes with $A$~=~112 and
$A$~=~124. These isotopes have been frequently chosen as targets and, more
recently, also as beams for isotopic studies (\cite{avde94,kunde96,xu99}
and references given in \cite{lozh92}). 

For the case of $^{112}$Sn and $^{124}$Sn targets
and, e.g., the production of $^7$Li fragments we may write 
\begin{equation}
\delta(^7{\rm Li})=
\frac{\sigma(^7{\rm Li};^{112}Sn)}{\sigma(^7{\rm Li};^{124}Sn)}=
\frac{\sigma(^7{\rm Li};N/Z=1.24)}{\sigma(^7{\rm Li};N/Z=1.48)}=
\delta(^7{\rm Li};\Delta(N/Z)=0.24).
\label{EQ3}
\end{equation}
This notation emphasizes the role of the $N/Z$ ratio of the source 
as the dominant parameter. It ignores the contribution of the projectile 
to the source which, however, may be justified for the very light 
projectiles (p, d, $\alpha$) 
that were used in the examples to be discussed in the following.
By defining a reduced isotopic effect
$\delta(X)/\delta(^6{\rm Li})$, i.e. the isotopic effect for species X
in relation to that for $^6$Li, uncertainties 
of the absolute normalizations of the measurements with the two
targets are eliminated. 
The reduced isotopic effect represents a double ratio which, e.g.,
for X = $^7$Li is 
equal to the ratio of the $^7$Li/$^6$Li yield ratios measured with the 
two targets. It is an observable equivalent to the value of 
the slope of the $^7$Li/$^6$Li yield ratio as a function of $N/Z$ 
(Fig. \ref{rati}).

\begin{figure}[t]
   \centerline{\epsfig{file=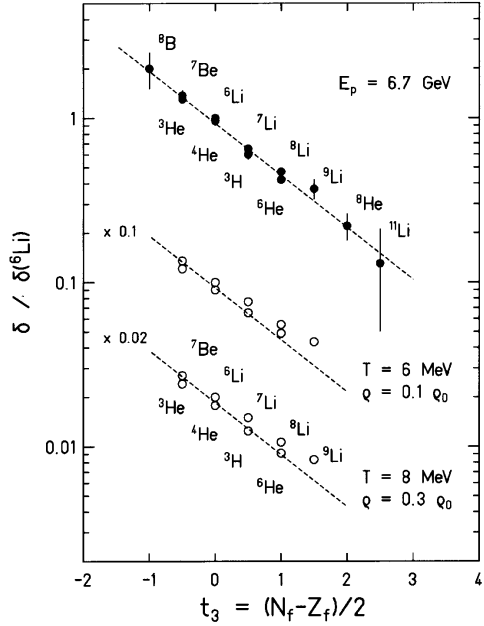,height=8.5cm}}
        \caption[]{\it\small
Reduced isotopic effect $\delta$/$\delta(^6{\rm Li})$ versus the third 
component of the isospin $t_3$. The experimental results for protons
of $E_p$ = 6.7 GeV \protect\cite{boga80} 
are given in the upper part (full circles). 
The dashed line represents an exponential fit to the data.
In the lower part the same fit curve is compared to the results obtained 
with the quantum-statistical model (open circles), calculated with two sets 
of parameters $T$ and $\rho$, and with $\Delta(N/Z)$ = 0.24
($\rho_0$ denotes the saturation density of nuclei,
from Ref. \protect\cite{lozh92}).
}
\label{isot}
\end{figure}

Figure 3 shows the reduced isotopic effects measured at the proton 
synchrotron of the JINR, Dubna, for the interaction of 
protons of $E_{\rm p}$ = 6.7 GeV 
with the target nuclei $^{112}$Sn and $^{124}$Sn
\cite{boga80}. These experiments were performed inclusively with 
multi-element telescopes positioned at approximately 90$^{\circ}$ with
respect to the beam. 
The cross section ratios differ by more than one order of 
magnitude for the neutron poor and the most neutron rich 
fragments. Plotted against the third component $t_3$ of the fragment 
isospin, the data exhibit a nearly perfect exponential dependence.
This result is a generalization of what was observed for the slopes 
of isotope ratios (Fig. \ref{rati}) and can also be 
understood on the basis of Eq. \ref{EQ2}.

Quantum-statistical models (QSM) are useful for extracting more 
quantitatively the information contained in 
measured isotope yield ratios or isotopic effects. Several versions of such 
models have been developed for different purposes 
\cite{hahn88,konop94,majka97,gulm97}. For the following comparison the
model of Hahn and St\"ocker \cite{hahn88} has been chosen.
It assumes thermal and chemical equilibrium at the breakup point where the 
fragmenting system is characterized by a density $\rho$, temperature $T$, 
and by its overall $N/Z$ ratio. The model respects fermion and boson 
statistics which, however, is not crucial at high temperature. It does not 
make provisions for the finite size of nuclear systems but follows the 
sequential decay of excited fragments according to tabulated 
branching ratios.

Results of QSM calculations are given in the lower half of Fig. \ref{isot}. 
They are compared to the experimental data represented by the 
dashed exponential fit curves. Pairs of temperatures $T$ and 
densities $\rho$ can apparently be found that give an accurate 
description of the observed $t_3$ dependence.
The temperatures chosen in these examples are in the range of 
measured breakup temperatures (see below), and the matching 
densities correspond to large breakup volumes 
as assumed in the statistical multifragmentation models
\cite{gross90,bond95}. Low densities in this range have been deduced
from proton-proton correlation functions measured for reactions in which
multifragmentation is dominant \cite{fritz99}. 
Their meaning in the light-ion induced reactions with small mean 
fragment multiplicities \cite{hsi97} is not fully clear at present.

\begin{figure}[t]
   \centerline{\epsfig{file=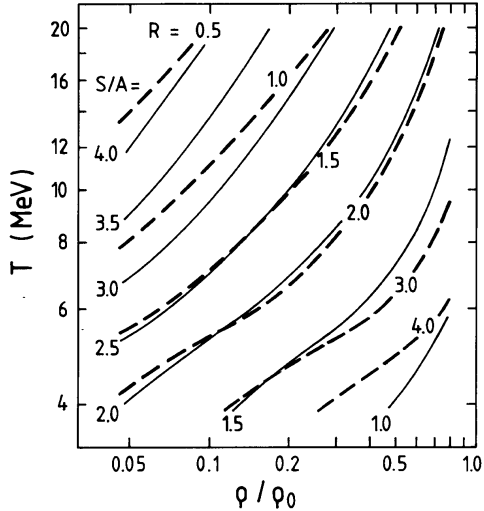,height=7cm}}
        \caption[]{\it\small
Results of calculations with the quantum-statistical model code
\protect\cite{hahn88} for the composite system $^{12}$C + $^{197}$Au.
The thin full lines are the isentropes in the temperature-versus-density
plane. The thick dashed lines are the contours of constant ratio $R$ of
the yields for $^7$Li and $^6$Li (from Ref. \protect\cite{traut87}).
}
\label{entr}
\end{figure}

The example illustrates that isotopic effects or isotopic yield ratios
in this way do not permit the individual determination of either the
breakup temperature or density but rather define a relation between them.
This was studied in more detail in Refs. \cite{wada87,traut87,tro88}
where it was found that the loci in the temperature-versus-density plane 
corresponding to fixed isotopic yield ratios effectively coincide
with isentropes (Fig. \ref{entr}). Obviously, 
the $N/Z$ ratio of the system has to be specified if this is to be used 
as a method to determine the entropy at breakup. 
Entropy values $S/A$
determined in this way for fragmentation channels in $^{12}$C and 
$^{18}$O induced reactions at bombarding energies up to 84 MeV per nucleon
are in the range of 1.5 to 2.5 \cite{wada87,tro88}.
More recent analyses based on similar methods 
have shown that this range of entropies is virtually an invariant over
a wide variety of fragmentation reactions \cite{lozh92,kuhn93}. 
The interest attracted by the entropy as an observable in heavy ion 
reactions is related to the approximately isentropic nature 
of the expansion phase \cite{bert83,papp95}. 
Entropy measurements thus permit a characterization of the excited 
systems formed during the initial impact and before expansion \cite{hahn88}.

\section{Equilibrium in Spectator Reactions}
\label{Sec_4}

In this section we will temporarily lift the restriction to isotopic 
degrees of freedom and more generally search for observables with the
potential of giving further evidence for equilibrium in fragmentation 
reactions. We will also proceed to a new type of reaction and 
investigate spectator decays initiated by collisions of heavy nuclei at 
relativistic bombarding energies \cite{dronten}.
These reactions are best viewed within the participant-spectator model
\cite{gosset} which distinguishes between the 'hot' 
hadronic system, formed by fast nucleons and secondary hadrons
produced in hard collisions, and the 'cold' spectators 
consisting of the remnants of the incident nuclei. It turns 
out that these spectators can also be quite highly excited, 
up to and beyond their total binding energies. 
Because of the nature of the excitation 
mechanism, predominantly nucleon knockout and absorption of slow 
nucleons recoiling from hard collisions, 
high degrees of equilibration may be expected.

We will focus on two recent experiments with the ALADIN 
spectrometer at the heavy-ion synchrotron SIS of the GSI Darmstadt
in which reactions of $^{197}$Au + $^{197}$Au in the regime of 
relativistic energies up to 1 GeV per nucleon were studied.
In the first experiment, the ALADIN spectrometer was used
to detect and identify the products of the projectile-spectator
decay \cite{schuet96}. The setup offers full coverage for 
projectile fragments with a dynamic range extending from helium isotopes 
to the original projectiles. Individual charge resolution and, 
for lighter fragments up to $A \approx$ 20, also individual mass 
resolution is obtained. The Large-Area Neutron Detector (LAND)
was used to measure coincident free neutrons emitted by the 
projectile source. In the second experiment, three multi-detector
hodoscopes, consisting of a total of 216 Si-CsI(Tl) telescopes,
and three high-resolution telescopes were positioned at backward angles
to measure the yields and correlations of isotopically resolved
light fragments of the target-spectator decay \cite{xi97}.
From these data excitation energies and masses, temperatures,
and densities were deduced. 

\begin{figure}[ttb]
   \centerline{\epsfig{file=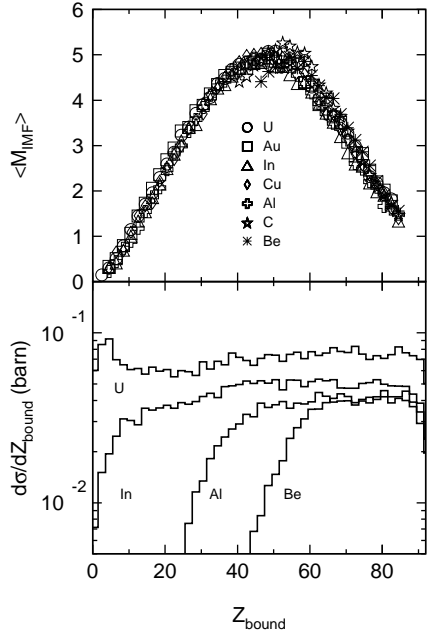,height=8.5cm}}
        \caption[]{\it\small
Top: Mean multiplicity of intermediate-mass
fragments $\langle M_{\rm IMF} \rangle$ as a function of
$Z_{\rm bound}$ for the reactions of $^{238}$U projectiles
at $E/A$ = 1000 MeV with the seven
targets Be, C, Al, Cu, In, Au, and U.
\protect\noindent
Bottom:
Measured cross sections $d\sigma /dZ_{\rm bound}$
for the reactions of $^{238}$U projectiles
at $E/A$ = 1000 MeV with the four targets
Be, Al, In, and U.
Note that the experimental trigger suppresses the very peripheral 
collisions at $Z_{\rm bound} \ge$ 70 which have much larger 
cross sections than indicated here
(from Ref. \protect\cite{schuet96}).
}
\label{mimf}
\end{figure}

For the presentation of these data, the quantity $Z_{\rm bound}$ 
has emerged as a useful sorting variable. $Z_{\rm bound}$ is equal
to the sum of the atomic numbers $Z_i$ of all projectile fragments
with $Z_i \geq$ 2. It reflects the variation of the charge of
the primary spectator system and is therefore correlated with the
impact parameter of the reaction. In the second type of experiments,
with high-resolution detectors looking at the target spectator,
the symmetric collision system $^{197}$Au + $^{197}$Au was studied.
$Z_{\rm bound}$ was determined for the projectile decay with the
time-of-flight wall of the ALADIN spectrometer, and it was assumed that 
its mean values for the target and projectile spectators 
are identical for a given event class.

\begin{figure}[ttb]
   \centerline{\epsfig{file=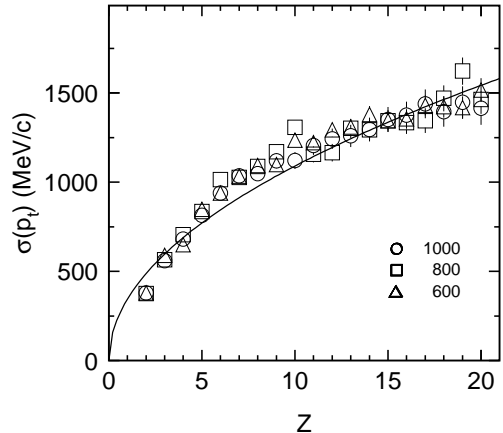,height=6cm}}
        \caption[]{\it\small
Widths of the transverse-momentum distributions
$\sigma (p_{\rm t})$ as a function of the fragment atomic number $Z$ 
for the reactions $^{197}$Au + $^{197}$Au at
$E/A$ = 600, 800, and 1000 MeV and for 20 $\le Z_{\rm bound} \le$ 60.
The line is proportional to $\sqrt{Z}$
(from Ref. \protect\cite{schuet96}).
}
\label{sigt}
\end{figure}

The universal features of the spectator decay,
as apparent in the
observed $Z_{\rm bound}$ scaling of the measured charge 
correlations, were the first and perhaps most striking indications 
for equilibrium \cite{schuet96}.
The target invariance of the $M_{\rm IMF}$ versus $Z_{\rm bound}$ 
correlation was first observed for collisions of $^{197}$Au projectiles
with C, Al, Cu, and Pb targets at 600 MeV per nucleon 
\cite{ogil91,kreutz}.
In Fig. \ref{mimf} (top) this correlation is shown 
for $^{238}$U projectiles at 1000 MeV per nucleon
and for a set of seven targets, ranging from Be to U.
The data for the lighter targets extend only over parts of the
$Z_{\rm bound}$ range, more clearly visible in the bottom 
part of the figure where the differential cross sections
$d\sigma/dZ_{\rm bound}$ for four out of the seven targets are displayed.
From the cross sections,
by assuming a monotonic relation between $Z_{\rm bound}$ and the
impact parameter, an empirical impact parameter scale can be obtained.
Central collisions lead to the smallest values of $Z_{\rm bound}$ that
can be reached with a given target, and given intervals of $Z_{\rm bound}$
may be reached with different targets but in collisions with
different impact parameters. The partitioning, apparently, depends only
on $Z_{\rm bound}$ and not on the impact parameter or target individually,
as long as they populate a given range of $Z_{\rm bound}$.

\begin{figure}[ttb]
   \centerline{\epsfig{file=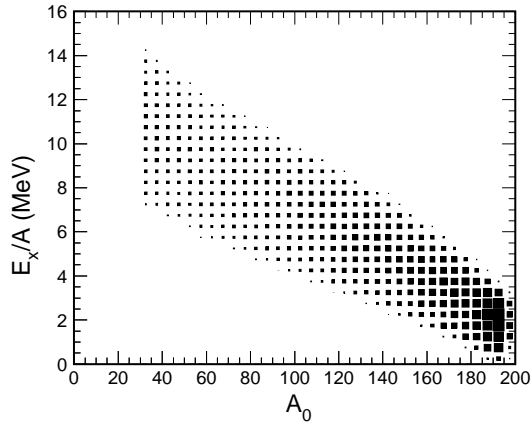,height=6cm}}
        \caption[]{\it\small
Ensemble of excited spectator nuclei used as input for the 
calculations with the statistical multifragmentation model
as a function of their excitation energy $E_x/A$ and mass $A_0$. 
The area of the squares is proportional to the intensity
(from Ref. \protect\cite{xi97}).
}
\label{ense}
\end{figure}

The invariance of the fragmentation patterns, when plotted as a function
of $Z_{\rm bound}$, suggests that the memory of the entrance channel 
and of the dynamics governing the primary interaction of the 
colliding nuclei is lost.
This extends to other observables; the transverse-momentum
widths of the fragments, as shown in Fig. \ref{sigt}, 
do not change with the bombarding energy,
indicating that collective contributions to the transverse motion are 
small. 
The equilibration of the three kinetic degrees of freedom in 
the moving frame of the projectile spectator was confirmed by the 
analysis of the measured velocity spectra \cite{schuet96}.
The square-root dependence on the atomic number $Z$ 
implies kinetic energies nearly independent of $Z$ and hence of the mass.

The success of the statistical multifragmentation models in describing
the observed population of the partition space may be seen as a
further argument for equilibration.
Here the main task consists of finding an appropriate ensemble of
excited nuclei to be subjected to the multi-fragment decay according to the 
model prescription. 
Starting from the entrance channel with models describing the primary phase 
of the collision may not necessarily provide sufficiently
realistic ensembles, even though a good description of the
fragment correlations was obtained with the 
quantum-molecular-dynamics model coupled
to the statistical multifragmentation model \cite{konop93}.
An alternative method consists of using empirical ensembles
derived by searching for an optimum reproduction of the observed
partitioning. Near perfect
descriptions of the measured correlations, including their dispersions
around the mean behaviour, can be achieved \cite{xi97,botv95}.
The mathematical procedure of backtracing allows for
studying the uniqueness of the obtained solutions and their
sensitivities to the observables that were used to generate 
them \cite{deses96}. As an example, the ensemble derived empirically
for the reaction $^{197}$Au + $^{197}$Au 
at 1000 MeV per nucleon is shown in Fig.~\ref{ense}. It extends over wide
ranges of mass and excitation energy with both quantities being
correlated as expected within the participant-spectator picture.

The spectator source, well localized in rapidity \cite{schuet96} 
and, apparently, exhibiting so many signs of equilibration, seems an
excellent candidate for studying the nuclear phase diagram. 
Limitations arise from the fact that nucleons and 
very light particles from the early reaction stages may appear at
spectator rapidities. Their contributions are difficult to suppress
or to identify which makes it difficult to extract
global variables such as the excitation energy of the system at 
the equilibrium stage. We will address this point in more detail 
when discussing the caloric curve of nuclei.

\section{Measurement of Temperature}
\label{Sec_5}

When measuring the temperature of excited nuclear systems one has 
to keep in mind that nuclei are closed systems with no external heat bath. 
Consequently, the temperature of the system cannot be pre-determined 
but has to be reconstructed from its decay products.
For a microcanonical ensemble, the thermodynamic temperature of a system 
may be uniquely defined in terms of the total-energy state density. 
An experimental determination of the state density 
and of its energy dependence is, however, hitherto impossible, 
except at very low excitations where this method is actually being used
\cite{melby99}. Nuclear temperature determinations, therefore, 
take recourse 
to 'simple' observables of specific degrees of freedom which 
constitute a good approximation to the true thermodynamic temperature.
In the canonical limit, the system may be seen as the heat bath for the 
chosen probe whose response is measured.

Several techniques have been developed for the determination of
temperatures of excited nuclear systems \cite{morri94}.
In the work leading to
the caloric curve of nuclei the method suggested
by Albergo {\it et al.} \cite{albergo} 
following earlier considerations of Hoyle \cite{hoyle46} has been
used. It is based on the assumption of chemical equilibrium and
requires the measurement of double ratios of isotopic yields.

In the limit of thermal and chemical equilibrium,
the double ratio $R$ built from the yields $Y_{\rm i}$ 
of two pairs of nuclides
with the same differences in neutron and proton numbers is given by 
\begin{equation}
R=\frac{Y_{1}/Y_{2}}{Y_{3}/Y_{4}} = 
a\cdot \exp(((B_{1}-B_{2})-(B_{3}-B_{4}))/T) =
a\cdot \exp(\Delta B/T)
\label{EQ4}
\end{equation}
where $B_{\rm i}$ denotes the binding energy of particle species i and the 
constant $a$ contains the mass numbers and internal partition sums. 
The chemical potential appearing in the expression for single isotope 
ratios (Eq. \ref{EQ2}) cancels in the appropriately chosen double ratio. 
The expression can therefore be solved with respect to the temperature $T$
provided the internal partition sums are known. The approximation usually 
made to circumvent this difficulty is to use the corresponding expression 
for the population of the ground states of the four isotopes, i.e. with the 
spin degeneracy factors for the ground states, and to treat 
modifications due to decays of higher-lying states as a perturbation. 
The problem of
sequential decay, presently one of the main limitations of temperature 
measurements, will be discussed in more detail below. 

A meaningful temperature scale can only be derived if the ratio $R$
is sufficiently sensitive to the temperature of the system. 
By differentiating Eq.~\ref{EQ4} we see that a relative error 
of the double yield ratio ${\Delta}R/R$ results in a
relative modification of the extracted temperature by
\begin{equation}
 \frac{{\Delta}T}{T} = - \frac{T}{{\Delta}B}\cdot\frac{{\Delta}R}{R}.
\label{EQ5}
\end{equation}
Thus, a stability of this thermometer against uncertainties of 
${\Delta}R/R$ will only result if the
binding energy difference $\Delta B=(B_1-B_2)-(B_3-B_4)$ is larger 
than the typical temperature to be measured. 
The analysis by Tsang {\it et al.} \cite{tsang97}
of more than 1000 possible isotope thermometers 
for the same reaction supports this rule of thumb.

\begin{figure}[ttb]
   \centerline{\epsfig{file=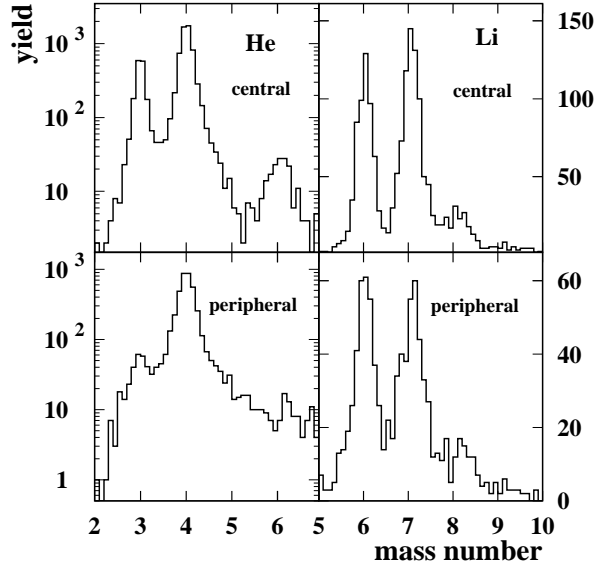,height=8cm}}
        \caption[]{\it\small
Mass spectra of He fragments (left panel)
and Li fragments (right panel) from the projectile spectator
following $^{197}$Au + $^{197}$Au collisions at $E/A$ = 600 MeV.
The upper and lower panels
correspond to central and peripheral collisions, respectively
(from Ref. \protect\cite{moeh95}).
}
\label{mass}
\end{figure}

Particularly large values for $\Delta B$ are obtained if a $^3He/^4He$ 
ratio is involved because the difference in binding energy of the 
two helium isotopes is 20.6 MeV.
It may be combined with, e.g., the lithium yield ratio 
$^{6}$Li/$^{7}$Li or with the hydrogen yield ratios p/d or d/t. 
Mass spectra for helium and lithium isotopes, measured for the
reaction $^{197}$Au + $^{197}$Au at $E/A$ = 600 MeV with the
ALADIN spectrometer, are shown in Fig. \ref{mass}. The 
strong variation of the $^{3}$He yields reflects the sensitivity of 
this less strongly bound nuclide to the variation of the
temperature with impact parameter.

Solving Eq. \ref{EQ4} for this case of $^3He/^4He$ and $^{6}$Li/$^{7}$Li
and in the ground-state approximation yields the following expression:
\begin{equation}
T_{\rm HeLi,0} = 13.3 MeV/\ln(2.2\frac{Y_{^{6}{\rm Li}}/Y_{^{7}{\rm Li}}}
{Y_{^{3}{\rm He}}/Y_{^{4}{\rm He}}}).
\label{EQ6}
\end{equation}

The subscript 0 of $T_{\rm HeLi,0}$ is meant to indicate that
Eq. \ref{EQ6} is strictly valid only for the ground-state 
population of the considered isotopes at the breakup stage.
The experimentally measured yields, however, contain all contributions 
from $\gamma$-unstable higher-lying states of the same isotope and 
from particle decays of other isotopes feeding the ground or 
$\gamma$-unstable states. 
In the work of the ALADIN collaboration \cite{poch95}, 
the expected magnitude of this effect was investigated by performing
calculations with the quantum-statistical model
\cite{hahn88,konop94}, a sequential-evaporation model \cite{char88},
and a statistical multifragmentation model \cite{gross86}. The primary
fragmentation process is treated very differently in these models but they
all follow explicitly the sequential decays of excited primary products.

\begin{figure}[ttb]
   \centerline{\epsfig{file=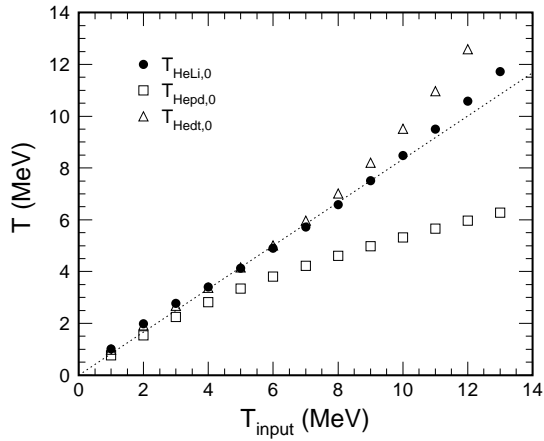,height=6cm}}
        \caption[]{\it\small
Temperatures $T_{\rm HeLi,0}$, $T_{\rm Hepd,0}$, and $T_{\rm Hedt,0}$, 
according to the quantum statistical model,
as a function of the input temperature $T_{\rm input}$. A breakup
density $\rho/\rho_0$ = 0.3 is assumed. The dotted line represents the 
linear relation $T_{\rm input}$/1.2
(from Ref. \protect\cite{xi97}).
}
\label{sequ}
\end{figure}

As an example, results obtained with the quantum statistical model 
are shown in Fig. \ref{sequ}. The assumed breakup density is
$\rho /\rho_0$ = 0.3 with $\rho_0$ denoting the saturation density.
Despite the strong feeding of the light particle yields through 
secondary decays, the relation between the input temperature 
of the model and $T_{\rm HeLi,0}$ as well as $T_{\rm Hedt,0}$ was 
found to be almost linear. Other temperature 
probes, as illustrated for $T_{\rm Hepd,0}$ in the figure, 
may be more strongly affected by sequential decays.
Variations of the input density within reasonable limits in this model
and the comparison with results obtained 
with the other decay models suggested that the accuracy of
these estimates may lie within $\pm$ 15\%. In order to account
for the systematic deviation from unity, a constant calibration 
factor $T_{\rm HeLi} = 1.2 \cdot T_{\rm HeLi,0}$ was adopted 
(corresponding to the dotted line in Fig. \ref{sequ}).

The consequences of sidefeeding from higher lying states have also
been investigated by other groups with different methods (see, 
e.g., Refs. \cite{majka97,gulm97,xi96,viola99} and references given therein).
The results differ considerably in magnitude as well as in the sign of the
required correction \cite{majka97,xi98c} and, in some cases, 
exceed the $\pm$15\% margin quoted above \cite{gulm97}. 
The treatment of the continuum part of the excitation spectrum of emitted 
fragments is conceptually and practically difficult \cite{gulm97,xi96}
and, furthermore, each isotope thermometer will require an
individual calibration (Fig. \ref{sequ} and Refs. \cite{kolo96,tsang96}). 
It is obvious that quantitative experimental 
information on the amount of sequential feeding will be needed, 
such as recently obtained for $^{129}$Xe + Sn reactions \cite{marie98},
and that experimental cross comparisons with 
alternative thermometers are mandatory in order to possibly
reduce the uncertainties of the isotopic temperature scale.

\begin{figure}[ttb]
   \centerline{\epsfig{file=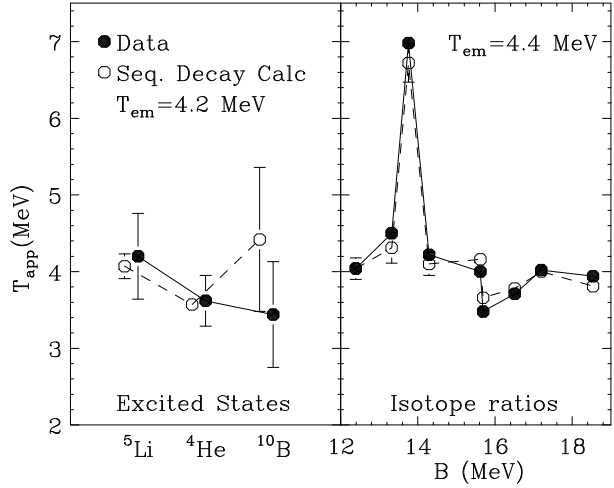,height=8cm,angle=90}}
        \caption[]{\it\small
Apparent temperatures obtained from relative populations of excited states 
for $^5$Li, $^4$He, and $^{10}$B nuclei (left panel) and from isotopic 
ratios (right panel, T$_{\rm HeLi,0}$ is given by the data point at
$B$ = 13.3 MeV). The closed points are the data and the open points 
are the predictions of sequential decay calculations
(from Ref. \protect\cite{huang97}).
}
\label{huang}
\end{figure}

A first cross comparison of the T$_{\rm HeLi}$ thermometer with 
excited-state temperatures deduced from excited state populations of
lithium, beryllium and boron isotopes for the $^{36}$Ar + $^{197}$Au 
reaction at 35 MeV per nucleon gave compatible results \cite{tsang96}. 
A similar consistency was also observed for central collisions of the 
heavier Au + Au system at the same energy \cite{huang97}. 
The quality of the agreement between the different thermometers 
investigated in the latter reaction is shown in Fig. \ref{huang}.
In the nomenclature used there, a distinction is made between the
'true' emission temperature at breakup $T_{\rm em}$ and the uncorrected
apparent temperature $T_{\rm app}$ obtained with a particular 
thermometer, such as $T_{\rm HeLi,0}$ in Eq. \ref{EQ6} \cite{emission}.
All temperature values, deduced either from relative populations of
states (left panel) or from isotopic yield ratios (right panel), 
are consistent with an emission temperature of 4.3 $\pm$ 0.1 MeV.
The figure also demonstrates that the sequential-decay calculations used 
for this case reproduce the observed behaviour of the apparent temperatures
rather well, including the large excursion of the apparent isotope 
temperature deduced from $^{3,4}$He and $^{9,10}$Be 
($\Delta B$ = 13.8 MeV). This anomaly has been traced down to the 
imbalance in the number of low-lying excited states in the $^{9,10}$Be
pair \cite{bond98}; it reflects the potentially dramatic influence 
of sequential decays in special cases. 

A cross calibration over a wider range of excitation energies
and reaction types was performed by studying central 
$^{197}$Au + $^{197}$A collision at 50 to 200 MeV per nucleon incident 
energy with the ALADIN spectrometer. For this purpose, the setup had been
supplemented by three hodoscopes and several high-resolution telescopes
positioned at angles close to mid-rapidity \cite{serf98}. 
The angular resolution and granularity of the hodoscopes
was optimized in order to permit the identification of excited particle 
unstable resonances in light fragments from the correlated detection of
their decay products. Yields of isotopically resolved light fragments
were measured with the individual telescopes which each consisted 
of three Si detectors of increasing thickness backed by a CsI(Tl) 
scintillation detector. 

\begin{figure}[tb]
   \centerline{\epsfig{file=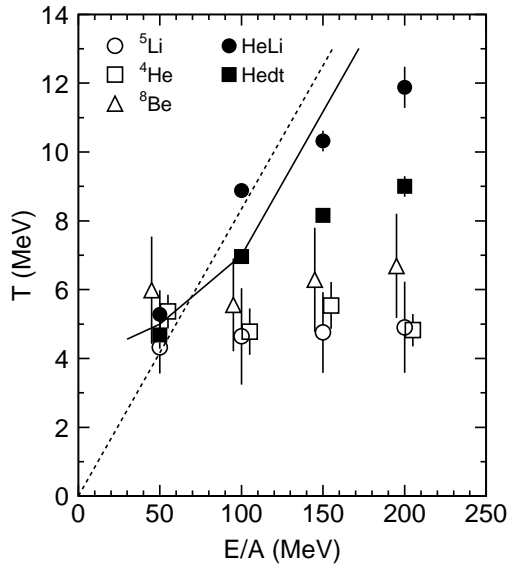,height=8cm}}
        \caption[]{\it\small
Measured isotope temperatures (full symbols) and excited-state temperatures
(open symbols) as a function of the incident energy per nucleon. The
indicated uncertainties are mainly of systematic origin. The meaning of the
lines is explained in the text (data from \protect\cite{serf98}).
}
\label{serf}
\end{figure}

The obtained values for two isotope and three
excited-state temperatures are given in Fig. \ref{serf}.
The isotope temperatures $T_{\rm HeLi}$ and $T_{\rm Hedt}$
were derived as described above, and a correction factor of 1.2 was 
applied in order to account for the effects of sequential feeding.
Excited-state temperatures were determined for $^4$He, $^5$Li, and 
$^8$Be, fragments with two widely separated states for which
modifications of the apparent emission temperatures due to feeding 
were expected to be small (Fig. \ref{huang} and 
Refs. \cite{poch87,schwarz93}). 

At 50 MeV per nucleon, all temperature values coincide within the interval
$T$ = 4 to 6 MeV in good agreement with the result of Huang {\it et al.}
\cite{huang97} obtained at 35 MeV per nucleon (Fig. \ref{huang}).
With rising beam energy, however, the isotope
temperatures rise linearly up to
$T_{{\rm HeLi}} \approx$ 12 MeV and $T_{{\rm Hedt}} \approx$ 9 MeV
at 200 MeV per nucleon (closed symbols).
The excited-state temperatures, on the other hand, appear to be 
virtually independent of the bombarding energy. They scatter closely
around their individual mean values of about 4.5 MeV up to 6 MeV 
for the three cases.
The same striking divergence of the two types of thermometers
has been made for central $^{86}$Kr + $^{93}$Nb reactions \cite{xi98b}
and for spectator reactions in $^{197}$Au + $^{197}$Au
at 1000 MeV per nucleon \cite{gross98}. 

Lacking at the moment a quantitative explanation of this surprising 
observation, it might be instructive to recall a similar phenomenon 
during the cosmic big-bang. Also there different degrees of freedom 
freeze out at various stages of the big-bang evolution,
hence signaling different temperatures. Of course, this cooling is 
intimately related to the existence of collective radial flow. 
It may, therefore, not be too surprising that the discrepancy between 
the two thermometers, in this case, emerges at incident energies exceeding 
50 MeV per nucleon at which radial collective flow starts to represent 
a significant part of the available collision energy \cite{reis97}. 
Rather crude estimates for the expected breakup temperatures if flow 
is taken into account are given by the lines in Fig. \ref{serf}. 
They qualitatively indicate the tendency exhibited by the isotope
temperatures. The interpretation of the excited-state temperatures
should also consider
that the resonances used for the temperature evaluation are very specific
quantum states with widths of 1 MeV or less. They are unlikely to
exist in the nuclear medium in identical
forms \cite{roep90,alm92,dani92}. In the case of volume breakup,
the asymptotic states that finally will be observed can develop and 
survive only at very low densities that may not be experienced by 
the cluster before it is effectively emitted into vacuum.
Excited-state temperatures may thus represent the internal fragment
temperatures at the very final stages of the fragmentation process.
The obtained mean value near 5 MeV corresponds well to results of 
dynamical calculations based on transport models \cite{fuchs97}.

Apparent temperatures deduced from purely thermal interpretations of the
Maxwellian-like kinetic-energy spectra are in 
clear disagreement with the isotope temperatures. There are many effects
that may influence the kinetic energies in these reactions, 
such as sequential decay, collective motion, Coulomb interaction,
preequilibrium emissions and Fermi motion. 
For spectator decays following $^{197}$Au + $^{197}$Au collisions, 
e.g., the kinetic temperatures are of the order of 15 to 20 MeV and much 
larger than the breakup temperatures discussed so far. It has recently 
been shown, however, that taking into account the effect of the nucleonic 
Fermi motion within the colliding nuclei will reconcile these seemingly 
contradictory observations \cite{gait00,odeh00}.

\section{A Caloric Curve of Nuclei}
\label{Sec_6}

Caloric curve is the term commonly adopted for a relation between
temperature $T$ and energy $E$. In thermodynamics, the caloric equation 
of state is given by the function $E(T)$ under well defined conditions
as, e.g., fixed volume and particle number in the simplest case.
Here we will speak of a caloric curve of nuclei in the looser sense in
that we use our experimental means to change the energy content of a 
nuclear system and simultaneously measure its temperature \cite{poch97}. 
The obvious motivation for this endeavour is
provided by the hope of establishing a link to the hypothetical
liquid-gas phase transition in extended nuclear matter \cite{hirsch99}.

At relativistic energies, only small fractions 
of the initial collision energy are converted into excitation energy of
the spectator. The energy deposition there can only be reconstructed from the 
exit-channel configuration which, however, requires a complete knowledge 
of all decay products with their atomic numbers, masses, and 
kinetic energies and including neutrons. Since this is not easy to obtain, 
at the least event by event, 
assumptions and approximations have to be made.

\begin{figure}[ttb]
   \centerline{\epsfig{file=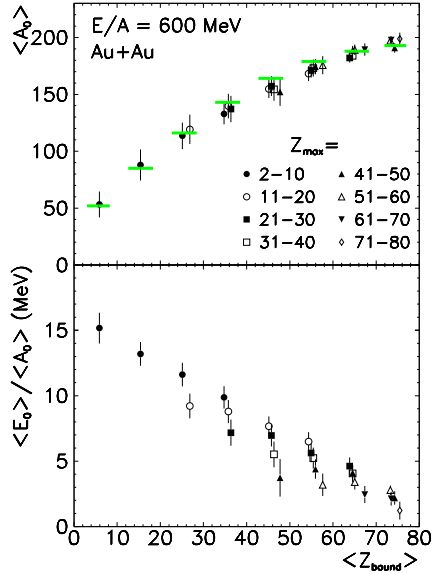,height=8cm}}
        \caption[]{\it\small
Reconstructed average mass $\langle A_0 \rangle$ (top) and excitation energy
$\langle E_0 \rangle / \langle A_0 \rangle$ (bottom) of the decaying
spectator system as functions of $Z_{\rm bound}$ and $Z_{\rm max}$.
The data symbols represent averages over 10-unit-wide 
bins in these two quantities as indicated.
The shaded horizontal bars (top panel) represent the
masses according to the participant-spectator model at the empirical
impact parameter deduced from $d\sigma /dZ_{\rm bound}$
(from Ref. \protect\cite{poch95}).
}
\label{exci}
\end{figure}

A method to determine the excitation energy from the experimental data
along this line was first presented by Campi {\it et al.} \cite{campi94}
and applied to the $^{197}$Au + Cu data measured by the ALADIN 
collaboration in their first experiment \cite{kreutz}. The caloric curve 
presented in Ref. \cite{poch95} was constructed from the more recent data 
for the $^{197}$Au + $^{197}$Au reaction at 600 MeV per nucleon.
Here also experimental information on neutron production, 
collected with LAND, was used. Hydrogen isotopes were not detected and
assumptions concerning the overall $N/Z$ ratio of the spectator,
the intensity ratios of protons, deuterons, and tritons, and the
kinetic energies of hydrogen isotopes had to be made. The
masses of the heavier fragments were not measured with sufficient 
precision and assumed to follow 
the EPAX parameterization \cite{suem90}.
The uncertainties resulting from the variation of these quantities within
reasonable limits were included in the errors assigned to the results.

The spectator masses $A_0$ and the specific excitation energies $E_0/A_0$, 
obtained from this analysis,
are given in Fig. \ref{exci} as functions of $Z_{\rm bound}$
and $Z_{\rm max}$, with $Z_{\rm max}$ denoting the
maximum atomic number detected within an event. 
The mean mass $\langle A_0 \rangle$ decreases with 
decreasing $Z_{\rm bound}$, in good agreement with the expectations 
from the geometric participant-spectator model \cite{gosset}.
Within a given bin of $Z_{\rm bound}$, $\langle A_0 \rangle$ 
is fairly independent of $Z_{\rm max}$.
The smallest mean spectator mass in the bin 
of $Z_{\rm bound} \le$ 10 is $\langle A_0 \rangle \approx$ 50.
The excitation energy $E_0$ appears to be a function 
of both $Z_{\rm bound}$ and $Z_{\rm max}$; 
events with smaller $Z_{\rm max}$, i.e. the
more complete disintegrations, correspond to the higher excitation 
energies.
The maximum number of fragments, observed at $Z_{\rm bound} \approx$ 40, 
is associated with initial excitation energies of
$\langle E_0 \rangle/\langle A_0 \rangle \approx$ 8 MeV.
With decreasing $Z_{\rm bound}$ the deduced excitation energies reach up
to $\langle E_0 \rangle/\langle A_0 \rangle \approx$ 16 MeV.

\begin{figure}[ttb]
   \centerline{\epsfig{file=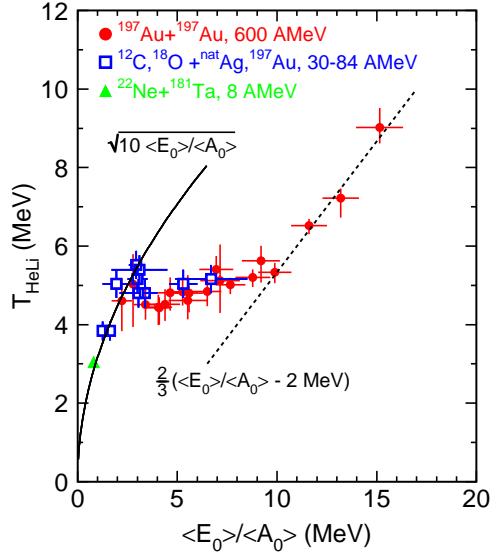,height=8cm}}
        \caption[]{\it\small
Caloric curve of nuclei as constituted by the
temperature T$_{\rm HeLi}$
as a function of the excitation energy per nucleon.
The lines are explained in the text
(from Ref. \protect\cite{poch95}).
}
\label{calo}
\end{figure}

The pairwise correlation of the isotope temperature, deduced as described 
in the last section, with the excitation energy leads to 
the caloric curve shown in Fig. \ref{calo}.
Besides the data from projectile decays following 
$^{197}$Au + $^{197}$Au collisions at 600 MeV per nucleon,
results from earlier experiments with $^{197}$Au targets at intermediate
energies 30 to 84 MeV per nucleon (c.f. Section \ref{Sec_2}) 
and for compound nuclei produced
in the $^{22}$Ne + $^{181}$Ta reaction are included. 
For the light-ion induced 
reactions at intermediate energies, the excitation energy
of the target residues was obtained by subtracting the energy lost in 
preequilibrium emissions from the collision energy while, in the 
compound case, it was assumed to be equal to the collision energy.

We first like to draw attention to the fact that the whole range of 
reaction channels from compound evaporation to near vaporization can be
covered with a single temperature observable.
The required helium and lithium isotopes 
are still produced in sufficient quantity at these extreme ends of the 
range of excitation energies. 
We further notice the consistency of the data obtained
for different types of reactions, suggesting that the smooth
S-shaped curve may represent a more general property of excited nuclei.
In fact, at low energies, the deduced temperatures
$T_{\rm HeLi}$ follow the sqare-root behaviour of a Fermi-liquid,
as represented by the full line calculated for 
a level density parameter $a = A/10$ MeV$^{-1}$.
At high excitation energies, the rise of the temperature seems to
approach a linear function with the slope 2/3 of a classical gas. 
In the limit of a free nucleon gas, the offset should be $\approx$ 8 MeV,
corresponding to the mean binding energy of nuclei.
A smaller offset may be caused by the finite density at freeze-out
and by the finite fraction
of bound clusters and fragments of intermediate mass 
that are present even at these high excitation energies. The offset
of 2 MeV is consistent with a breakup density 
$\rho /\rho_0$ between 0.15 and 0.3 \cite{poch95}.

\begin{figure}[ttb]
   \centerline{\epsfig{file=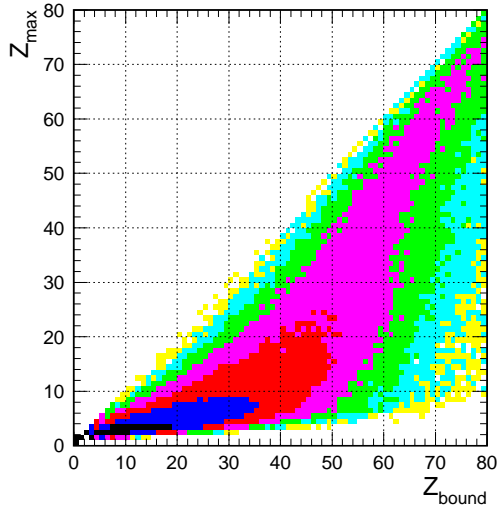,height=7cm}}
        \caption[]{\it\small
$Z_{\rm max}$-versus-$Z_{\rm bound}$ distribution measured 
in the reaction $^{197}$Au + $^{197}$Au at 1000 MeV per nucleon.
Conventional fission events are removed, the
shadings follow a logarithmic scale
(from Ref. \protect\cite{odeh99}).
}
\label{zmzb}
\end{figure}

Within the range of $\langle E_0\rangle /\langle A_0\rangle$ from
3 MeV to 10 MeV, the temperature increases very little while the system 
passes from the 'liquid' regime, governed by the degrees of freedom of a 
heavy residue, to that of the 'vapor' of light fragments and particles.
This transition in the dominant characteristics of the reaction channel
is illustrated in Fig. \ref{zmzb} which shows the event distribution in the 
$Z_{\rm max}$-versus-$Z_{\rm bound}$ plane. At large $Z_{\rm bound}$,
corresponding to small excitation energies, 
the ridge line starts at $Z_{\rm max} \approx Z_{\rm bound}$, i.e.
with events containing only one large fragment or heavy residue.
The most probable $Z_{\rm max}$ then drops fairly rapidly 
near $Z_{\rm bound} \approx$ 50 and approaches very small values 
at the smallest $Z_{\rm bound}$.
The observation of 
$Z_{\rm max} \ll Z_{\rm bound}$ in this region of large 
excitation energies implies that the system has disintegrated into 
a larger number of smaller fragments. The energy
needed for the formation of smaller constituents, correspondingly,
limits the rise of the temperature. 

The qualitative interpretation of the caloric curve along these lines is 
motivated by its reminiscence of what we expect for a first-order phase 
transition. It has to be refined and possibly confirmed with model studies. 
Statistical multifragmentation models have been used to identify the 
individual contributions to the total excitation energy and to 
quantify the origin of the second rise at $E_x/A \approx$ 10 MeV
\cite{bond95,radu99}. It was found that the internal degrees of freedom
of fragments of intermediate mass are still quite important at these high 
excitations \cite{bond95}. Transport theoretical approaches, 
based on antisymmetrized molecular-dynamics models, give the possibility to 
study nuclear systems at finite pressure \cite{schnack97,suga99}. 
Caloric curves obtained in this way in model experiments include a rise
at high excitations that has been shown to correspond to a nucleon gas with 
Van-der-Waals properties \cite{suga99}.

Alternative interpretations, in particular for the 
temperature rise at the highest energies, have been presented by several
groups. In the interpretation of 
Natowitz {\it et al.} \cite{nato95}, the variation of the
system mass (Fig. \ref{exci}) is seen as the primary reason for the
observed variation of the temperature and related 
to the mass dependence of the temperatures limiting the stability
of excited nuclei \cite{bonche}.
In the expansion scenario modeled by Papp and 
N\"orenberg \cite{papp95}, the temperatures
in the plateau region are found to be consistent with a spinodal
decomposition in the dynamically unstable region of the 
temperature-versus-density plane. The upbend at high excitation energies,
in this model, indicates a concentration of the deduced 
breakup densities at a minimum value around $\rho /\rho_0 \approx$ 0.3. 

Another line of interpretation emphasizes the time dependence of the 
emission process. The time evolution 
has been studied with the expanding-emitting
source model which assumes statistical emissions from
an expanding and thereby cooling source \cite{fried90}. 
According to these calculations,
the less bound nuclei are more likely to be emitted at earlier 
reaction times, with the effect that integrated yield ratios 
represent a complex convolution of the temperature profile and 
the emission probabilities as a function of the reaction time \cite{viola99}.
Temperatures obtained from integrated yield ratios, 
consequently, have the character
of weighted mean values over finite distributions. Differential 
analyses have been performed in
order to extract cooling curves \cite{xi98a} or to obtain
temperature values for fixed reaction times \cite{cibor00}.
For some reactions a systematic difference of the kinetic energy spectra of 
the $^{3,4}$He nuclei has been observed,
suggesting that the two isotopes originate from different reaction
stages \cite{boug99,neub00}. It is therefore desirable to extend
the range of temperature observables to other combinations of isotopes
that do not include the helium nuclei. The $T_{\rm BeLi}$ temperature,
derived from the $^{7,9}$Be and $^{6,8}$Li isotope ratios appears
to be a promising candidate; 
the second rise at high excitation energies
is also seen with this observable \cite{muell99}.

\begin{figure}[ttb]
   \centerline{\epsfig{file=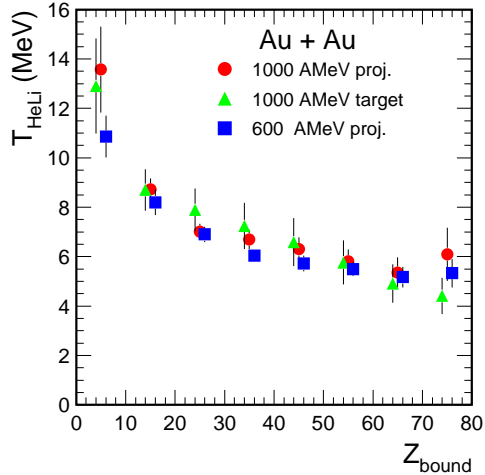,height=6.5cm}}
        \caption[]{\it\small
Temperatures $T_{\rm HeLi}$ for target ($E/A$ = 1000 MeV)
and projectile spectators ($E/A$~=~600 and 1000 MeV)
as a function of $Z_{\rm bound}$. The data symbols represent averages 
integrated over $Z_{\rm max}$ and over 10-unit wide 
bins of $Z_{\rm bound}$ (from Ref. \protect\cite{odeh99}).
}
\label{tinv}
\end{figure}

The discussion clearly shows that asking for a final interpretation of the 
caloric curve of nuclei would be premature at the present time. Furthermore,
ongoing analyses and new data have also led to small modifications of the 
originally published curve. As a result of previously not included
experimental information and corrections, the temperatures deduced 
for projectile fragments in the $^{197}$Au + $^{197}$Au reaction 
at 600 MeV per nucleon have increased by between 10\% and 20\% \cite{xi97}. 
The new temperature values $T_{{\rm HeLi}}$ are shown in Fig. \ref{tinv}
as a function of $Z_{\rm bound}$ and, unlike Fig. \ref{calo}, averaged 
over $Z_{\rm max}$. They are compared to 
the results obtained at 1000 MeV per nucleon, for both the
projectile and target spectators. For the target spectator,
isotopic yields were measured at $\theta_{\rm lab} = 150^{\circ}$ with
a four-element telescope while $Z_{\rm bound}$ was determined for the
coincident projectile decay with the ALADIN time-of-flight wall.
The agreement with the temperatures obtained for the projectile
decay at the same energy is expected from the symmetry of the collision
and only illustrates the accuracy achieved in these experiments.
The coinciding results for 600 and 1000 MeV per nucleon incident energy,
on the other hand, indicate an invariance with respect to the bombarding 
energy that is well known from other observables (Section \ref{Sec_4}).
Calculations with the statistical multifragmentation 
model, using the ensemble of excited spectator nuclei 
shown in Fig. \ref{ense}, are in good agreement with these measurements
\cite{xi97}. The simultaneous reproduction of 
the observed charge partitions and of this temperature-sensitive 
observable represents a necessary requirement
for a consistent statistical description of the spectator fragmentation.

More recent data and analyses have also shown 
that the excitation energies
obtained from calorimetry do not exhibit a similar invariance with 
respect to the incident energy. It was found that the mean kinetic
energies of neutrons, 
measured at three bombarding energies with LAND, increase considerably
with the bombarding energy \cite{gross99}. Together with the contribution
of the protons, this leads to a difference of about 40\% in the 
reconstructed excitation energy for the collisions at 
600 and 1000 MeV per nucleon. 
Most likely, the experimental nucleon yields contain contributions 
from earlier reaction stages even if they are measured in 
the kinematic regimes corresponding to spectator decays. This may also
provide the explanation for the excess of the experimentally determined 
energies over those obtained from backtracing with the 
statistical multifragmentation model 
(Figs. \ref{ense}, \ref{exci}, and
Ref. \cite{schuet96}). Detailed studies of light particle emissions in 
these reactions, including excitation functions, will be needed in order to 
disentangle the various sources of light particles. At present,
this uncertainty prevents the construction of a fully invariant 
caloric curve from spectator decays.

With the same or similar techniques caloric curves have been derived 
by other groups for other 
types of reactions \cite{cibor00,ma97,haug98,kwiat98,dago99}. 
The transition 
from the liquid branch to a plateau-like behaviour, coinciding with 
the onset of fragment production, 
seems to be a general feature of all of these curves. The second rise,
however, is not seen as pronounced as for the 
$^{197}$Au + $^{197}$Au spectator reactions (Figs. \ref{calo},\ref{tinv}).
With lighter projectiles, the cross sections for reaching the high 
excitation energies at which this rise occurs are rather small 
(cf. Fig. \ref{mimf}). On the other hand, the vaporization of light systems
into hydrogen and helium isotopes, 
as observed for the $^{36}$Ar + $^{58}$Ni reaction at 95 MeV per nucleon
\cite{bord99}, has been described with a chemical-equilibrium model and 
temperatures of up to 24 MeV \cite{gulm97}.

\section{Concluding Remarks}
\label{Sec_7}

This chapter has been meant to illustrate the usefulness of isotopic yield 
ratios for a statistical description of heavy-ion reactions. The main 
topics were the questions of how to obtain evidence for equilibration 
and of how to actually measure thermodynamical variables. 
From there a very natural path has 
led to phase transitions in finite nuclei, a topic of highest
current interest. 
It is obvious that none of these
subjects has been covered exhaustively, and we refer the
reader to the cited literature and to the relevant articles collected in 
this volume.

The underlying concept of equilibrated breakup states necessarily 
constitutes an idealization in view of the rapid dynamic evolution of 
energetic reaction processes. Its limitations were evident in the 
discussion of multifragment emissions, most notably in the 
comparison of the different temperature observables. The dominant role
of phase space, on the other hand, is apparent from the successful 
description of a variety of global characteristics of these reactions
that is possible with the assumption of a statistical breakup.

The increasing availability of secondary beams for nuclear reactions
will allow future studies of isospin degrees of freedom to be extended 
over a wider range of neutron-to-proton asymmetries. 
The two-fluid nature of nuclear matter promises new phenomena to be seen
in experiments with asymmetric nuclei,
as discussed elsewhere in this volume. 
Isotopic yield 
ratios can be expected to remain in the center of interest as
important observables probing the source composition 
as well as its thermodynamical properties. 
It is also not excluded that spectator reactions
at high energies will prove unique for producing 
globally equilibrated asymmetric systems over the desired
wide range of excitation energies. 

{\it The authors would like to thank their colleagues of the ALADIN 
collaboration and at the GSI for support and discussions during the 
preparation of this manuscript.}


\begin{thebibliography}{99}
\bibitem{emission}
The term emission temperature is sometimes exclusively used 
for temperatures extracted from relative populations 
of excited states, see e.g. Ref. \cite{poch97}
\end{thebibliography}
\end{document}